\documentstyle[tighten,preprint,prc,aps,epsfig]{revtex}
\draft
\newcommand {\nc} {\newcommand}
\nc {\beq} {\begin{eqnarray}}
\nc {\eeq} {\end{eqnarray}}
\nc {\eeqn} [1] {\label{#1} \end{eqnarray}}
\nc {\eoln} [1] {\label{#1} \\}
\nc {\eol} {\nonumber \\}
\nc {\ve} [1] {\mbox{\boldmath $#1$}}
\nc {\la} {\mbox{$\langle$}}
\nc {\ra} {\mbox{$\rangle$}}
\nc {\cL} {\mbox{${\cal L}$}}
\nc {\dem} {\mbox{$\frac{1}{2}$}}
\begin {document}
\title{Equivalence of the Siegert-pseudostate and 
Lagrange-mesh $R$-matrix methods}
\author{
D.~Baye, J.~Goldbeter, and J.-M.~Sparenberg \\
}
\address{
Physique Nucl\'{e}aire Th\'{e}orique et Physique
Math\'{e}matique, C.P. 229, \\ 
Universit\'{e} Libre de Bruxelles, B 1050 Brussels, Belgium
}
\date{\today}
\maketitle
\begin{abstract}
Siegert pseudostates are purely outgoing states at some 
fixed point expanded over a finite basis. 
With discretized variables, they provide an accurate description of 
scattering in the $s$ wave for short-range potentials with few basis states. 
The $R$-matrix method combined with a Lagrange basis, 
i.e.\ functions which vanish at all points of a mesh but one, 
leads to simple mesh-like equations which also allow an accurate 
description of scattering. 
These methods are shown to be exactly equivalent for any 
basis size, with or without discretization. 
The comparison of their assumptions shows how to accurately 
derive poles of the scattering matrix in the $R$-matrix formalism 
and suggests how to extend the Siegert-pseudostate method to 
higher partial waves. 
The different concepts are illustrated with the Bargmann potential 
and with the centrifugal potential. 
A simplification of the $R$-matrix treatment can usefully be extended 
to the Siegert-pseudostate method. 
\end{abstract}
\clearpage
\noindent
\section{Introduction}
Recently a new approach for the description of scattering 
using Siegert pseudostates has been proposed 
\cite{TON97,TON98}. 
Siegert states are bounded solutions of the Schr\"odinger 
equation which are purely outgoing at infinity \cite{Si-39}. 
These states are particularly interesting because their 
complex wave numbers provide the poles of the scattering matrix. 
However the derivation of these states and their use is not 
computationally simple. 
Therefore, Tolstikhin, Ostrovsky, and Nakamura \cite{TON97,TON98}
have proposed to use modified Siegert states which satisfy purely 
outgoing conditions at some finite distance. 
These modified states correspond to exact Siegert states 
of a truncated potential. 
The cutoff distance is similar to the channel radius in the $R$-matrix 
theory and its introduction opens a way to possible relations 
between the two methods. 

The Siegert pseudostates are defined by an expansion of the Siegert 
states of the truncated potential over a finite basis which becomes 
complete when the number of basis states tends to infinity. 
In Refs.~\cite{TON97,TON98}, the authors present an efficient way 
for deriving the Siegert pseudostates in the $s$ wave 
for short-range potentials. 
They establish a number of remarkable mathematical properties 
of these pseudostates and of the corresponding complex wave numbers.
This method can be simplified with the help of a Gauss quadrature 
in the spirit of the discrete-variable representation \cite{LHL85}. 
Then a simple matrix representation is obtained 
where the potential matrix is diagonal. 

The $R$-matrix theory is a powerful tool, not only to parametrize 
scattering matrices and cross sections but also to solve the 
Schr\"odinger equation at positive energies 
\cite{LT-58,BRT83,BD-89}. 
In this method, the configuration space is divided in two parts, 
separated at the channel radius. 
In the external part, the wave functions are approximated 
by their asymptotic expressions. 
In the internal part, a finite basis of square-integrable functions 
can be used. 

The Lagrange-mesh method is an approximate variational 
calculation which resembles a mesh calculation \cite{BH-86,VMB93,BHV02}. 
This property is obtained by using a basis of Lagrange functions, 
i.e.\ orthonormal functions which vanish 
at all points of an associated mesh but one, 
and the Gauss quadrature corresponding to this mesh. 
In spite of its simplicity, the accuracy of the Lagrange-mesh method 
is very high, a property not explained yet \cite{BHV02}. 

In the single-channel case, Malegat \cite{Ma-94} combined 
the $R$-matrix theory with a Lagrange-mesh method based on 
shifted Legendre polynomials to study the scattering by a simple 
solvable potential. 
Strikingly, the accuracy of the $R$-matrix method on a mesh is as good 
as the accuracy of the $R$-matrix method using the corresponding 
Lagrange basis without any approximation \cite{BHS98}. 
This method can easily be extended to multichannel scattering 
and gives accurate results for realistic problems \cite{HSV98}.

In the present paper, we show that the Siegert-pseudostate method 
and the $R$-matrix method on a Lagrange mesh are completely 
equivalent for any basis size when the bases used in both approaches 
are identical. 
More strikingly both methods remain exactly equivalent 
when their respective mesh approximations are employed.   
This equivalence sheds new lights on both approaches. 
It emphasizes the poorly known fact that the $R$-matrix method 
can give a direct access to the poles of the scattering matrix. 
The technique presented in Refs.~\cite{TON97,TON98} provides 
an accurate practical way of solving this problem for the 
$s$ wave of short-range potentials. 
Symetrically, the validity of the $R$-matrix method for higher 
partial waves and for long-range potentials indicates the way to 
natural generalizations of the Siegert-pseudostate method. 

The $R$-matrix method is summarized in Sec.~II. 
Its application to the Lagrange-Legendre mesh is presented in Sec.~III. 
The equivalence with the Siegert-pseudostate method 
is discussed in Sec.~IV. 
Examples are commented in Sec.~V.
Concluding remarks are presented in Sec.~VI. 

\section{$R$-matrix method}
As in Ref.~\cite{TON98}, we restrict ourselves to a single 
channel.
Contrary to that reference, we first consider an arbitrary partial 
wave and potentials with a possible Coulomb asymptotic behavior. 
We follow the notations of Ref.~\cite{BHS98}. 
A translation into the notations of Ref.~\cite{TON98} is delayed 
to Sec.~IV (see also Table \ref{tab:1} in that section). 

For the $l$th partial wave, the radial Schr\"odinger equation 
can be written as
\beq
(H_l - E)u_l=0
\eeqn{2.1}
with $u_l (0) = 0$. 
With $\hbar = m = 1$, the radial Hamiltonian reads 
\beq
H_l = T_l + V(r) 
= \frac{1}{2} \left(-\frac{d^2}{dr^2} + \frac{l(l+1)}{r^2}\right) + V(r),
\eeqn{2.2}
where $V(r)$ is a radial potential. 
The phase shift $\delta_{l}$ is obtained from the asymptotic behavior 
of bounded solutions 
\beq
u_l(r) \mathop{\longrightarrow}\limits_{r \rightarrow \infty} 
I_l (k r) - S_{l} O_l (k r),
\eeqn{2.3}
for positive values of the wave number $k$ corresponding 
to the energy $E=\dem k^2$. 
The functions $I_{l}$ and $O_{l}$ are ingoing and outgoing Coulomb 
functions and $S_l = \exp(2i\delta_l)$ is the scattering matrix. 
In the following $k$ will also take complex values. 

In the $R$-matrix method, the configuration space is divided at the channel 
radius $a$ into an internal region and an external region. 
In the external region, the wave function is approximated by its 
asymptotic form (\ref{2.3}). 
In the internal region, it is expanded on some basis. 
The formalism is conveniently expressed with the help of the 
Bloch surface operator \cite{Bl-57}
\beq
\cL(B) = \frac{1}{2}\, \delta (r-a) 
\left( \frac{d}{dr} - \frac{B}{r} \right),
\eeqn{2.4}
where $B$ is the boundary parameter. 
The Bloch-Schr\"odinger equation reads
\beq
(H_l +\cL(B) -E) u_l = \cL(B) u_l,
\eeqn{2.5}
where the operator $H_l +\cL(B)$ is Hermitian when $B$ is real. 
The approximation consists in using the asymptotic form ({\ref{2.3}) 
in the right-hand side of Eq.~({\ref{2.5}). 
The main advantage of the $R$-matrix method is that an expansion 
in square-integrable functions can be used in the left-hand side. 

Let us consider a set of $N$ basis functions $f_i (r)$ 
(not necessarily orthogonal) and let us expand $u_l$ 
in the internal region as 
\beq
u_l (r) = \sum_{j=1}^N c_j f_j (r).
\eeqn{2.6}
Equation (\ref{2.5}) becomes after projection on $f_i (r)$, 
\beq
\sum_{j=1}^N \, [C_{ij}(B) -E N_{ij}] c_j 
= \la f_i | \cL (B) | I_l - S_l O_l \ra.
\eeqn{2.7}
The elements of the symmetric matrices $\ve{C}$ and $\ve{N}$ are defined as
\beq
C_{ij}(B) & = & \la f_i | T_l + \cL(B) + V | f_j \ra
\eol
& = & C_{ij}(0) - (B/2a) f_i (a) f_j (a)
\eeqn{2.8}
and 
\beq
N_{ij} = \la f_i | f_j \ra.
\eeqn{2.8a}
They correspond to one-dimensional integrals over the variable $r$ 
from 0 to $a$. 
Matrix $\ve{N}$ reduces to the unit matrix when the basis is orthonormal. 
The right-hand side of Eq.~(\ref{2.7}) is even simpler. 
Because of the Bloch operator, it only involves values at $r=a$. 

The coefficients $c_j$ are obtained by solving Eq.~(\ref{2.7}). 
The continuity condition at $r=a$ between the internal approximation 
(\ref{2.6}) and the asymptotic expression (\ref{2.3}) reads 
\beq
\sum_{j=1}^N c_j f_j (a) = I_l (k a) - S_{l} O_l (k a).
\eeqn{2.9}
Let us introduce the external logarithmic derivative $L_l$ 
at the channel radius 
\beq
L_l = ka \frac{O'_l(ka)}{O_l(ka)}
\eeqn{2.11}
and the dimensionless $R$ matrix 
\beq
R_l (B) = (2a)^{-1} 
\sum_{i,j=1}^N f_i(a) [\ve{C}(B)-E\ve{N}]^{-1}_{ij} f_j(a).
\eeqn{2.12}
The dependence of the $R$ matrix on the energy $E$ is implied. 
Introducing the $c_j$ in Eq.~(\ref{2.9}) and using 
Eqs.~(\ref{2.12}) and (\ref{2.11}), 
one obtains the scattering matrix for the $l$th partial wave 
\beq
S_l = \frac{I_l (ka)}{O_l (ka)} \ 
\frac{1 - (L_l^* - B) R_l (B)}{1 - (L_l - B) R_l (B)},
\eeqn{2.10}
where $L_l^*$ is the conjugate of $L_l$. 
Expression (\ref{2.10}) has the striking property that it 
does not depend on the boundary parameter $B$, independently 
of the size of the basis. 
Indeed, from the matrix relation (\ref{A.3}) in Appendix A, one obtains 
\beq
\frac{1}{R_l(0)} = \frac{1}{R_l(B)} + B
\eeqn{2.13}
for any $B$, real or complex. 
Introducing relation (\ref{2.13}) in Eq.~(\ref{2.10}) shows 
that any $B$ value leads to the same scattering matrix as for $B=0$. 
Equation (\ref{2.13}) is well known in $R$-matrix theory 
(see Eq.~(IV.2.5a) of Ref.~\cite{LT-58}). 
However its validity for the approximation (\ref{2.12}) 
for any basis size \cite{LR-69} is sometimes overlooked. 

The wave function in the internal region is then given by 
\beq
u_l (r) = [2a R_l (B)]^{-1} [I_l (ka) -S_l O_l (ka)] 
\eol \times \sum_{j=1}^N f_j (r) 
\sum_{i=1}^N [\ve{C}(B)-E\ve{N}]^{-1}_{ij} f_i(a).
\eeqn{2.13a}
As the scattering matrix $S_l$ and the external wave function 
$I_l (kr) -S_l O_l (kr)$, this expression does not depend on the 
choice for $B$. 
Indeed, with the help of relation (\ref{A.3a}), 
one easily shows that, for any $B$, it is equal to the similar 
expression where $B$ is replaced by zero. 

In Refs.~\cite{BHS98,HSV98}, the parameter $B$ was chosen 
equal to zero for obvious reasons of simplicity. 
Another interesting choice is \cite{Bl-57,LR-69} 
\beq
B = L_l.
\eeqn{2.14}
This complex value leads to a complex function $R_l (L_l)$ 
which is not an $R$ matrix in the strict sense since 
$R$ matrices are real. 
However it is also given by expression (\ref{2.12}). 
Equation (\ref{2.10}) then takes the simpler form \cite{LR-69} 
\beq
S_l = e^{-2i \phi_l (ka)} \ [1 + 2i P_l(ka) R_l (L_l)],
\eeqn{2.15}
where $P_l(ka)$ is defined as 
\beq
P_l = \mbox{$\frac{1}{2i}$} (L_l - L_l^*)
\eeqn{2.16}
and $\phi_l (ka)$ is half the phase of $O_l (ka)/I_l (ka)$. 
When $k$ is real, $P_l$ is the penetration factor 
given by the imaginary part of $L_l$ 
and $\phi_l (ka)$ is the hard-sphere phase shift \cite{LT-58}. 
Since Eq.~(\ref{2.15}) has no denominator, a direct 
relation appears between the poles of the scattering matrix 
and of the complex $R$ matrix. 
\section{$R$ matrix on a Lagrange mesh}
The previous section is valid for arbitrary bases. 
The calculation of the elements of matrix $\ve{C}$, 
\beq
C_{ij}(B) = \la f_i | T_l + \cL(0) | f_j \ra 
- (B/2a) f_i(a) f_j(a) + \la f_i | V | f_j \ra,
\eeqn{3.1}
involves an evaluation of the matrix elements of the potential, 
which can be tedious and must be repeated when the potential changes. 
By chosing a Lagrange basis and using the associated Gauss quadrature, 
one can avoid this calculation without losing accuracy \cite{BHS98}. 

As Lagrange basis, we use functions based on Legendre polynomials 
\cite{BH-86,Ma-94,BHS98,HSV98}. 
These functions are denoted as ${\hat f}_i$ in Ref.~\cite{BHS98}. 
Here as in Ref.~\cite{HSV98} we drop the `hat' 
because we shall not use any other basis. 
A Lagrange basis is a set of $N$ functions $f_i(x)$ associated with 
a Lagrange mesh of $N$ points $ax_i$ on the interval $[0,a]$ 
\cite{BH-86,VMB93}. 
The $x_i$'s are zeros of the shifted Legendre polynomial $P_N (2x-1)$ 
\cite{Ma-94}, i.e., 
\beq
P_N (2x_i-1) = 0.
\eeqn{3.2}
The Lagrange functions are continuous and indefinitely differentiable 
anywhere. 
They read 
\beq 
f_i(r) = (-1)^{N-i} a^{-1/2} \sqrt{\frac{1-x_i}{x_i}}\ 
\frac{r P_N[2(r/a)-1]}{r-ax_i}. 
\eeqn{3.3}
They satisfy the Lagrange conditions 
\beq
f_i (ax_j) = (a\lambda_i)^{-1/2} \delta_{ij},
\eeqn{3.4}
i.e., each $f_i$ vanishes at all mesh points $ax_j$, 
except at $ax_i$. 
The coefficients $\lambda_i$ are the weights associated with a 
Gauss-Legendre quadrature approximation for the $[0,1]$ interval. 
The Gauss quadrature on the $[0,a]$ interval reads \cite{Sz-67}
\beq
\int_0^a g(r) \, dr \approx a \sum^N_{k=1} \lambda_k g(a x_k).
\eeqn{3.5}
The weights $\lambda_i$ are equal to the traditional Gauss-Legendre weights 
for the $[-1,+1]$ interval, divided by 2. 

The Lagrange functions (\ref{3.3}) are not orthogonal \cite{BHS98}
\beq
\la f_i | f_j \ra = \delta_{ij} + (-1)^{i+j} \frac{1}{2N + 1} 
\sqrt \frac{(1-x_i) (1-x_j)}{x_i x_j}.
\eeqn{3.6}
Because of the Lagrange conditions (\ref{3.4}), they are approximately 
orthogonal at the Gauss approximation (\ref{3.5}), 
\beq
\la f_i | f_j \ra \stackrel{\rm Gauss}{=} \delta_{ij}.
\eeqn{3.7}
Strikingly, using the Gauss approximation does not seem to reduce the 
accuracy of the $R$-matrix method \cite{BHS98}. 

At the Gauss approximation, the potential matrix 
\beq
\langle f_i | V | f_j \rangle \stackrel{\rm Gauss}{=} V(ax_i) \delta_{ij} 
\eeqn{3.8}
is diagonal because of Eq.~(\ref{3.4}), and easy to compute. 
The other matrix elements are {\it exactly} calculated with the Gauss 
quadrature. 
The matrix elements of the sum of the radial part of the kinetic energy 
and of the Bloch operator are given by 
\beq 
\la f_i | T_0 + \cL(0) | f_i \ra = 
\frac{1}{6a^2x_i(1-x_i)} \left[ 4N(N+1)+3 + \frac{1-6x_i}{x_i(1-x_i)} \right]
\eeqn{3.9}
and, for $i \ne j$,
\beq 
\la f_i | T_0 + \cL(0) | f_j \ra = 
\frac{(-1)^{i+j}}{2a^2 [x_i x_j (1-x_i) (1-x_j)]^{1/2}} 
\nonumber \\ \times 
\left[ N(N+1)+1 + \frac{x_i+x_j-2x_ix_j}{(x_i-x_j)^2} 
- \frac{1}{1-x_i} - \frac{1}{1-x_j} \right].
\eeqn{3.10}
Finally, the remaining necessary expressions read 
\beq
\la f_i | r^{-2} | f_j \ra = a^{-2} x_i^{-2} \delta_{ij} 
\eeqn{3.11}
and
\beq
\la f_i | r^{-1} \delta(r-a) | f_j \ra = a^{-2} (-1)^{i+j} 
[x_i x_j (1-x_i) (1-x_j)]^{-1/2}.
\eeqn{3.12}
\section{Equivalence of the Siegert-pseudostate and $R$-matrix methods}
\subsection{The Siegert-pseudostate method}
First we briefly summarize the Siegert-pseudostate method. 
For this, we keep the $R$-matrix notations introduced above. 
Also, we start with definitions for an arbitrary partial wave. 
To avoid confusions, the equations of Ref.~\cite{TON98} are denoted 
as the reference number followed by the equation number in that reference. 

The Siegert pseudostates $\phi_l^{(n)} (r)$ with complex wave numbers 
$k_l^{(n)}$ are solutions of the equation 
\beq
(H_l +\cL(L_l^{(n)}) - E_l^{(n)}) \phi_l^{(n)} = 0,
\eeqn{4.1}
where $L_l^{(n)}$ is calculated for $k = k_l^{(n)}$ 
and $E_l^{(n)} = \dem k_l^{(n)2}$. 
Indeed, with the choice $B=L_l$, the right-hand side of Eq.~(\ref{2.7}) 
vanishes for purely outgoing waves such as the Siegert pseudostates. 
For $l=0$ and short-range potentials, 
definition (\ref{4.1}) is exactly equivalent to the pair of equations 
\cite{TON98}-(1a) and \cite{TON98}-(1c') (see Ref.~\cite{Bl-57}). 
Thanks to the use of the Bloch operator, Eq.~(\ref{4.1}) is more compact. 

Expanding $\phi_l^{(n)}$ in the internal region as 
\beq
\phi_l^{(n)} (r) = \sum_{j=1}^N c_j^{(n)} f_j (r),
\eeqn{4.2}
one obtains the homogeneous part of Eq.~(\ref{2.7}), 
\beq
\sum_{j=1}^N \, [C_{ij}(L_l^{(n)}) - E_l^{(n)} N_{ij}] c_j^{(n)} = 0.
\eeqn{4.3}
However, the dependence of this equation on its eigenvalues $k_l^{(n)}$ 
is strongly non linear because $k$ appears not only in $E=\dem k^2$ 
but also in $L_l$. 

We now specialize to $l=0$ and short-range potentials. 
Then the free outgoing wave is given by $O_0 = \exp(ikr)$ 
and the logarithmic derivative reads
\beq
L_0 (ka) = ika.
\eeqn{4.4}
With the help of Eq.~(\ref{2.8}), the system (\ref{4.3}) can be written as
\beq
\sum_{j=1}^N \, [C_{ij}(0) - \dem ik_0^{(n)} f_i (a) f_j(a) 
- \dem k_0^{(n)2} N_{ij}] c_j^{(n)} = 0.
\eeqn{4.5}
The eigenvalues $k_0^{(n)}$ appear linearly and quadratically 
because of the simple form (\ref{4.4}) of $L_0$. 
Ref.~\cite{TON98} provides an efficient algebraic algorithm 
for solving the system (\ref{4.5}), which leads to $2N$ eigenstates. 
The quadratic matrix eigenvalue problem is replaced by a standard 
generalized eigenvalue problem of double size. 
Such a simple algorithm is not yet available for system (\ref{4.3}) 
with $l > 0$. 

The fact that Eq.~(\ref{4.5}) can be solved algebraically 
gives an access to the physical poles of the scattering matrix 
related to the true Siegert states. 
The other obtained poles are either Siegert states of the truncated 
potential or non-converged Siegert states (see example below). 
Tolstikhin, Ostrovsky, and Nakamura have shown [Eq.~\cite{TON98}-(59)] 
that the approximate scattering matrix can then be written 
as a sum on poles under the form  
\beq
S_0 (k) = e^{-2ika} \left[ 1 + ik \sum_{n=1}^{2N} 
\frac{[\phi_0^{(n)} (a)]^2}{k_0^{(n)}(k_0^{(n)} -k)} \right].
\eeqn{4.6a}
Equivalently, a product expression for the scattering matrix 
reads [Eq.~\cite{TON98}-(61)] 
\beq
S_0 (k) = e^{-2ika} \prod_{n=1}^{2N} \frac{k_0^{(n)} +k}{k_0^{(n)} -k}.
\eeqn{4.6}
This elegant result is valid only for the $s$ wave. 

The internal wave function ($r \le a$) is given by equation 
\cite{TON98}-(57) which reads in the present notations 
\beq
u_0 (r) = -ik e^{-2ika} \sum_{n=1}^{2N} 
\frac{\phi_0^{(n)}(r) \phi_0^{(n)} (a)}{k_0^{(n)}(k_0^{(n)} -k)}.
\eeqn{4.7}
Equations (\ref{4.6a}) and (\ref{4.7}) assume that the Siegert 
pseudostates $\phi_0^{(n)}$ are properly normalized 
[see Eq.~\cite{TON98}-(28)]. 
\subsection{Equivalence for identical bases}
The above equations now allow us to prove the equivalence between 
the $R$-matrix technique of Refs.~\cite{Ma-94,BHS98} and the 
Siegert-pseudostate method of Refs.~\cite{TON97,TON98}. 
We shall first show that the approximations giving the scattering matrix 
are identical for any common finite basis without mesh approximation. 
To this end, we specialize to $l=0$ and to short-range 
potentials. 

In Refs.~\cite{Ma-94,BHS98}, the boundary parameter $B$ is taken 
equal to zero but, as proved in Sec.~III, exactly the same results 
would be obtained with any other value. 
Therefore we now focus on the choice $B=L_0$ with the $s$-wave 
logarithmic derivative (\ref{4.4}). 
The penetration factor $P_0$ and the hard-sphere phase shift $\phi_0$ 
take the simple forms 
\beq
P_0 (ka) = ka
\eeqn{4.8}
and 
\beq
\phi_0 (ka) = ka.
\eeqn{4.9}
Hence, Eq.~(\ref{2.15}) reads 
\beq
S_0 (k) = e^{-2i ka} \ [1 + 2i ka R_0 (ika)].
\eeqn{4.10}
This expression has the same structure as Eq.\ \cite{TON98}-(58), 
since the $R$ matrix is known to be related to the Green function through 
\beq
R_l = (2a)^{-1} G_l (a,a),
\eeqn{4.11}
see Eq.~(IV.1.10) of Ref.~\cite{LT-58}. 
It is thus equivalent to Eq.~(\ref{4.6}) and 
relates the $S$-matrix and complex $R$-matrix poles.  

In order to prove the equivalence of both methods, we have to 
compare the approximate calculations of these expressions 
for finite bases. 
In Ref.~\cite{TON98}, the Green function is obtained with 
Eq.~\cite{TON98}-(49), where the matrix is obtained by inversion 
from Eq.~\cite{TON98}-(50). 
Since the matrix appearing in Eq.~\cite{TON98}-(50) 
is identical to the present matrix $\ve{C}(L_0)-E\ve{N}$ 
[see Eq.~(\ref{2.8})], 
the equivalence with our expression (\ref{2.12}) is proved. 
\subsection{Equivalence of the mesh treatments}
We have just shown that, with the same finite basis, both methods 
are exactly equivalent. 
Now we show that the same property holds for the 
Discrete-Variable-Representation (DVR) approximation \cite{LHL85} 
of Ref.~\cite{TON98} and the Lagrange-mesh approximation of 
Ref.~\cite{BHS98} which is summarized in section III. 
For $l=0$, both methods make use of a mesh approximation related to zeros 
of Legendre polynomials. 

The $R$ matrix is given by Eq.~(\ref{2.12}) with matrix $\ve{C}$ 
calculated with expression (\ref{3.1}) where the different terms 
can be obtained on the mesh from Eqs.~(\ref{3.6}) and (\ref{3.8}) 
- (\ref{3.12}). 
In fact the DVR approximation for the Green-function matrix in Ref.~\cite{TON98} 
is the inverse of a matrix which is proportional to the Lagrange-mesh 
approximation of matrix $\ve{C}(ika) - E \ve{N}$ [Eq.~(\ref{3.1})]. 

The relations between the present quantities and those of 
Ref.~\cite{TON98} are detailed in Table \ref{tab:1}. 
The first line of the Table shows that the notation $x_i$ represents 
different zeros in both papers: 
in Ref.~\cite{TON98}, they are zeros of a standard Legendre polynomial 
$P_N (x)$ and belong to $[-1,1]$ while here and in Refs.~\cite{Ma-94,BHS98} 
they are zeros of a shifted Legendre polynomial $P_N (2x-1)$ 
and belong to $[0,1]$. 
This is only a notational difference. 
The second line shows the connection between the Gauss weights. 
The factor of two arises from the different lengths 
of the intervals. 
The basis functions are related in the third line. 
The functions $\pi_i(x)$ in Ref.~\cite{TON98} are chosen in such 
a way that they provide a representation of the unit operator 
[Eqs.~\cite{TON98}-(5) and \cite{TON98}-(6)] (see also Ref.~\cite{We-00}). 
The property \cite{TON98}-(C13) of these functions shows that 
they verify a Lagrange condition. 
A compact expression for them has already been derived in Ref.~\cite{BH-86} 
and used in Refs.~\cite{Ma-94,BHS98} [see the present Eq.~(\ref{3.3})]. 
The present functions $f_i (r)$ may seem to differ by a factor 
$r = \dem a (x+1)$ but, in Ref.~\cite{TON98}, 
this factor is included in the operator as in Ref.~\cite{Ma-94}. 
The overlaps $\rho_{ij}$ differ from the present $N_{ij}$ by a simple 
factor. 
Notice that the present form (\ref{3.6}) is simpler than expression 
\cite{TON98}-(C22) because we do not employ two different bases. 

Since the potential matrix elements are approximated in the same way, 
only the equivalence of the treatments of the kinetic energy remains 
to be proved. 
Let us detail the derivation of expressions (\ref{3.9}) and (\ref{3.10}). 
Up to a factor $1/2$, these matrix elements become 
with the Gauss quadrature 
\beq
- \int_0^a f_i(r) f''_j(r) dr 
+ \int_0^a f_i(r) \delta (r-a) f'_j(r) dr 
\nonumber \\
= - (a\lambda_i)^{1/2} f''_j (ax_i) + f_i (a) f'_j (a).
\eeqn{4.12}
The Gauss quadrature is exact for polynomials up to degree $2N-1$ 
\cite{Sz-67} so that expression (\ref{4.12}) is exact and leads to 
Eqs.~(\ref{3.9}) and (\ref{3.10}) 
(see Ref.~\cite{BHS98} for technical details). 
Equivalently, Eq.~(\ref{4.12}) can be written as 
\beq
\int_0^a f'_i(r) f'_j(r) dr  
= a \sum_{k=1}^N \lambda_k f'_i (ax_k) f'_j (ax_k).
\eeqn{4.13}
This expression is also exact but less compact. 
Still another approach is used in Ref.~\cite{TON98}: 
the left-hand side of Eq.~(\ref{4.13}) is evaluated 
analytically after expressing the functions $f_i$ in the basis of Legendre 
polynomials, in the spirit of the DVR method \cite{LHL85}. 
The notations for the kinetic-energy matrix elements are compared 
in the fifth line of Table \ref{tab:1}. 
Comparing Eqs.~\cite{TON98}-(C20) and \cite{TON98}-(C21) 
for $\tilde{K}_{ij}$ with the present Eqs.~(\ref{3.9}) and (\ref{3.10}) 
shows that the kinetic-energy matrix elements are calculated much more 
easily in the Lagrange-mesh philosophy than in the DVR philosophy. 
We have checked numerically that the expression \cite{TON98}-(C20) 
provides exactly the same results as ours, as it should. 
The notations for the full matrices are compared in the last line 
of Table \ref{tab:1}. 

Finally, let us mention a difference between the practical applications 
of the methods of Ref.~\cite{TON98} and of Refs.~\cite{Ma-94,BHS98}. 
In Refs.~\cite{Ma-94,BHS98}, expression (\ref{3.6}) 
has been replaced by its Gauss approximation (\ref{3.7}): 
the overlap matrix $\ve{N}$ is replaced by the unit matrix. 
This simplification is not used in Ref.~\cite{TON98}. 
In Ref.~\cite{TON98}, the algorithm requires that the equivalent of matrix 
$\ve{C}(ika)$ be multiplied to the left and to the right by 
$\ve{N}^{-1/2}$. 
This can easily be done in the present framework (see Sec.V) 
with $\ve{N}^{-1/2}$ calculated as explained in Appendix A. 
However this complication is useless at the practical level 
because it does not improve the accuracy \cite{BHS98}. 
We shall come back on the interest of the simplification (\ref{3.7}) 
when dealing with the first example in Sec.~V. 
\subsection{Consequences}
After those lengthy but necessary technical considerations, 
let us try to learn some practical consequences from the equivalence 
of the methods. 

For $s$ states, the equivalence of both methods provides a new approach 
to the determination of the poles of the $S$ matrix in the $R$-matrix 
formalism. 
Indeed deriving the complex $S$-matrix poles from the real $R$-matrix poles 
is not obvious. 
However the Siegert pseudostates $\phi_l^{(n)}$ appear naturally 
in the $R$-matrix formalism. 
They are solutions of Eq.~(\ref{4.1}) with complex wave numbers $k_l^{(n)}$. 
For $l=0$, the algebraic algorithm of Ref.~\cite{TON98} provides 
an efficient way of determining some poles of the $S$-matrix 
with sufficient accuracy. 
Note that only a few physical poles need usually be determined 
since the $S$ matrix is more conveniently given by Eq.~(\ref{2.10}) 
than by Eqs.~(\ref{4.6a}) or (\ref{4.6}). 

Symmetrically the equivalence is also useful to attack the same 
problem for higher partial waves. 
The authors of Ref.~\cite{TON98} have tried without success 
to generalize their search of Siegert pseudostates to $l>0$. 
They make use of Jacobi polynomials adapted to the value of $l$. 
This basis was shown in Ref.~\cite{BHS98} to be not more efficient 
than basis (\ref{3.3}), but more complicated to use. 
The reason for this failure is not due to a technical choice of basis 
but seems rather rooted in the technique of calculation of the 
scattering matrix. 
Indeed, as illustrated later in the second example below, 
the approximate {\it wave functions} determined in Ref.~\cite{TON98} 
should be accurate, up to a normalization factor. 
As shown at the end of Sec.~II, the same approximate wave functions 
are obtained for any choice of boundary parameter, 
independently of its physical adequacy. 
This includes the complex choice (\ref{4.4}) 
which is implicit in Ref.~\cite{TON98}. 
Hence only the scattering matrix is inaccurate in Ref.~\cite{TON98}. 

The natural generalization of the Siegert-pseudostate method 
to $l > 0$ and long-range potentials is Eq.~(\ref{4.3}). 
However this equation does not allow to use the algebraic 
technique because the non linearity is not any more quadratic. 
We think that it would be useful to use Eq.~(\ref{4.3}) 
anyway to derive physical poles of the $S$ matrix. 
The fact that it seems hopeless to find in this way all 
the pseudostates is of little importance since the $S$ matrix 
can easily and accurately be calculated with the $R$-matrix 
equation (\ref{2.10}). 
The search for the physical poles could for example be performed by 
extending the iterative algorithm of Descouvemont and Vincke \cite{DV-90}. 
\section{Illustrative examples}
\subsection{Bargmann potential}
Many examples are treated in Ref.~\cite{TON98} and we have reproduced 
these results. 
In the case of phase shifts, we have checked that we obtain the same values 
both with the Siegert-pseudostate method and with the $R$-matrix method 
within the accuracies of both numerical algorithms. 
When the number $N$ of mesh points is not large enough, 
both methods provide essentially identical inaccurate results. 
Rather than repeating here one of those examples, 
we have chosen a different one which provides interesting 
intuitive information on the notion of Siegert pseudostate. 
We also use this example to discuss more deeply the effect 
of the Gauss approximation (\ref{3.7}) on the overlap 
matrix element (\ref{3.6}). 

The Bargmann potential \cite{Ba-49,CS-77} is defined as
\beq
V(r) = -4b^2 \beta \frac{e^{-2br}}{(1+\beta e^{-2br})^2}
\eeqn{5.1}
with $\beta = (b-c)/(b+c)$ where $b$ and $c$ are real parameters. 
This potential has the remarkable property that its Jost function 
has only one pole and one zero 
\beq
f_0(k) = \frac{k+ic}{k+ib}.
\eeqn{5.2}
The potential has one bound state for $c<0$ or one virtual state for $c>0$. 
The scattering matrix reads 
\beq
S_0(k) = \frac{f_0 (-k)}{f_0 (k)} = \frac{(k+ib)(k-ic)}{(k-ib)(k+ic)}.
\eeqn{5.3}
It possesses the symmetry property $S_0 (bc/k) = S_0 (k)$. 
How approximation (\ref{4.6}) simulates expression (\ref{5.3}) 
is instructive. 

The wave functions of the Siegert states of the potential truncated 
at $a$ read
\beq
\phi_0^{(n)} (r) \propto \sin k_0^{(n)}r 
+ \frac{b^2-c^2}{k_0^{(n) 2}+b^2}
\frac{k_0^{(n)} \tanh br \cos k_0^{(n)}r - b \sin k_0^{(n)}r}
{b + c \tanh br},
\eeqn{5.4}
where $k_0^{(n)}$ is a solution of the equation
\beq
\frac{b^2 (c + ik)}{\cosh^2 ba} \sin ka 
= k \left[ (b^2 + ikc) \tanh^2 ka + b(c+ik) \tanh ba \right.
\eol
\left. - \frac{k^2+b^2}{b^2-c^2} (b + c \tanh ba)^2 \right] \exp(-i ka).
\eeqn{5.5}
The unique Siegert state of the Bargmann potential (\ref{5.1}) 
is given by Eq.~(\ref{5.4}) with the wave number $-ic$ 
which is an approximate solution of (\ref{5.5}) when $a$ is large. 

The values $b = 2$ and $c = -1$ are selected, which lead to a single 
bound state at energy $- 1/2$. 
Exact wave numbers of the Bargmann potential truncated at $a=5$, 
i.e.\ replaced by zero beyond that value, are displayed as dots 
in Fig.~\ref{fig:1} and in the first two columns of Table \ref{tab:2}. 
The isolated wave number close to $i$ corresponds to the bound state. 
It is not exactly $i$ because of the finite value of $a$. 
The pole of the simple Jost function (\ref{5.2}) is simulated by a line 
of wave numbers with imaginary parts close to $-2i$ and rather regularly 
spaced real parts. 
The accuracies of the different approximations can be checked 
with respect to these exact values. 

In a first step, we perform a calculation of the wave numbers 
corresponding to the Siegert pseudostates by following the technique 
of Ref.~\cite{TON98} without further approximations. 
For $a = 5$ and $N = 25$, the obtained values are depicted 
as circles in Fig.~\ref{fig:1}. 
For real parts comprised between about $-6.5$ and $6.5$, 
they closely correspond to the exact wave numbers of the cutoff 
Bargmann potential [see panel (a) of Fig.~\ref{fig:1}]. 
The remaining poles split from the exact ones. 
They do not have any physical significance. 
They correspond to unconverged solutions of Eq.~(\ref{4.5}) 
as shown by the comparison with the exact values in Fig.~\ref{fig:1}. 
When $N$ is increased, the horizontal line of physical poles 
of the cutoff potential obtained with the Siegert-pseudostate 
method extends to larger values and the non physical branches 
move accordingly. 

Numerical values are provided in Table \ref{tab:2} for $N = 25$. 
The first ones agree within better than $10^{-5}$ with the 
exact values. 
These results are unstable: tiny changes in the computational algorithm 
(or a change of computer) modify the non converged digits. 
This instability appears to be related to factors $1 - x_i$ in the 
denominators of expressions (\ref{3.9}), (\ref{3.10}), and 
(\ref{3.12}), for zeros $x_i$ close to 1. 
We have also observed that the accuracy on the first wave 
numbers progressively deteriorates when $N$ becomes larger than 25. 
This behavior is due to the fact that the norms of the corresponding 
eigenvectors are very small so that few digits of the eigenvalues 
are significant. 
When $N$ increases beyond some value, the number of significant 
digits of each wave number decreases but the number of physically 
significant wave numbers nevertheless increases as mentioned above. 
Because of this important instability of the results for large $N$, 
the value $N=25$ is close to optimal for $a=5$ in the present case. 

Now we add a simplification to the method by replacing 
the overlap matrix $\ve{N}$ by a unit matrix according 
to the Gauss approximation (\ref{3.7}). 
This approximation has been used in the $R$-matrix 
treatment of Refs.~\cite{BHS98,HSV98}, 
but it does not introduce much simplification in that case. 
The algorithm of Ref.~\cite{TON98} is much more simplified by this 
approximation. 
The obtained wave numbers are depicted as crosses in Fig.~\ref{fig:1}. 
All the physical poles of the cutoff potential remain 
essentially unmodified (see details in Table \ref{tab:2}). 
On the contrary, the structure of the unphysical poles 
is completely different [see panel (b) of Fig.~\ref{fig:1}]. 
As we shall see, this does not affect the phase shift 
at low energies. 

The first values obtained with approximation (\ref{3.7}) 
are essentially identical to those without that approximation 
(see Table \ref{tab:2}). 
This is very surprising because the agreement is much better 
than the accuracy on these values. 
When the real part increases, the values become more different, 
as already illustrated by the figure, 
but the results obtained with the Gauss approximation first remain 
better than those obtained with the exact overlap matrix 
[see panel (a) of Fig.~\ref{fig:1}]. 
This effect is also observed in other applications of the 
Lagrange-mesh method \cite{BHV02}. 

Phase shifts calculated for different energies with Eq.~(\ref{4.6}) 
are displayed in Table \ref{tab:3} and compared 
with the exact values from Eq.~(\ref{5.3}). 
Because of the symmetry property of the $S$ matrix, 
the phase shifts at energies $E$ and $b^2 c^2/4E$ add to $\pi$. 
The calculations are performed for two choices of channel 
radius $a$, i.e.\ 5 and 6, and for two numbers of mesh points, 
i.e.\ 25 and 40. 

For $a = 5$ and $N = 25$, the relative accuracy on the phase 
shift is about $10^{-7}$, which is better than for the wave numbers.
We have also calculated the phase shifts with Eq.~(\ref{4.6a}). 
The results are inaccurate as mentioned in Ref.~\cite{TON98} 
because the normalization of the Siegert pseudostates is difficult 
to achieve numerically: both terms of the normalization formula 
\cite{TON98}-(26) nearly cancel each other for several states. 
In the following we only refer to Eq.~(\ref{4.6}). 
Essentially the same values are obtained with the $R$-matrix mesh 
method and with both variants of the Siegert-pseudostate method. 
A difference only appears at $E = 10$ but is not significant. 
The three results are much closer to each other than to the 
exact one. 
Increasing the number of mesh points to $N = 40$ without modifying 
$a$ does not really improve the situation at $E = 0.1$ but is more 
useful at higher energies. 
The remaining disagreement is due to the value of $a$ 
for which the potential is not yet fully negligible. 
 
For $a = 6$ and $N = 25$, the relative accuracy on the phase 
shift is only about $10^{-5}$. 
The number of mesh points is too small for this $a$ value. 
At all energies, the three approximate results are much closer 
to each other than to the exact phase shift. 
When $N$ is increased to 40, the relative accuracy on the phase 
shifts is better than $10^{-9}$. 
No significant differences between the methods appear. 
The influence of the approximation (\ref{3.7}) is weak for 
all the considered energies. 
\subsection{Purely centrifugal potential}
In Ref.~\cite{TON98}, the Siegert-pseudostate method is applied 
to a purely centrifugal potential. 
In this case, the natural choice (\ref{2.14}) for $B$ would be 
\beq
L_1 = - 1 + \frac{(ka)^2}{1-ika},
\eeqn{5.6}
since $O_1(kr) = -i(1-1/ikr) \exp(ikr)$. 
The choice $B = L_0 = ika$ is made instead in order to use 
the algebraic algorithm established for $l=0$ short-range potentials. 
The $S$ matrix is then calculated with Eq.~(\ref{4.11}) which leads 
to very inaccurate results as can be expected in the present framework. 

The corresponding wave function is however essentially independent 
of the particular choice for $B$ (see Sec.~II). 
Hence, it is as accurate for any non-physical value of $B$ 
as for the physical one (\ref{5.6}). 
However, its normalization is not correct. 
We have verified that the method of Ref.~\cite{TON98} and in 
particular Eq.~\cite{TON98}-(57) or the present Eq.~(\ref{4.7}) 
provide accurate free-particle wave functions for the 
centrifugal potential, up to a multiplicative factor. 
This fact seems to have remained unnoticed by the authors 
of Ref.~\cite{TON98}. 
It can be understood from the present Eq.~(\ref{2.13a}) 
to which Eq.~(\ref{4.7}) is equivalent. 
Expression (\ref{2.13a}) is valid for any $B$ but the correct 
normalization factor $I_l (ka) -S_l O_l (ka)$ is not available 
to the Siegert-pseudostate method since $L_0$ is used instead 
of $L_1$ and the calculated $S$ matrix is inaccurate. 
Moreover, this correct scattering wave function is obtained 
with unphysical Siegert pseudostates and $S$ matrix poles. 
These unphysical states are just used as a basis for the 
expansion of the wave function. 

Our result has been obtained with basis (\ref{3.3}) based on 
Legendre polynomials without recourse to the more complicated 
Jacobi polynomials employed in Ref.~\cite{TON98}. 
This confirms the fact established in Ref.~\cite{BHS98} 
that the $l$-dependent basis of Jacobi polynomials, although accurate, 
is not necessary to treat $l>0$ partial waves. 
\section{Conclusions}
In this paper, we have proved that the Siegert-pseudostate and 
Lagrange-mesh $R$-matrix methods are in fact exactly equivalent. 
This property is true when the same finite basis is used in both 
approaches but also when the mesh methods described in Ref.~\cite{TON98} 
and in Refs.~\cite{BHS98,HSV98} are employed. 

This equivalence provides an approximate way of calculating the poles 
of the $S$ matrix in the $R$-matrix framework for $s$ wave short-range 
potentials. 
It also shows how to generalize the determination of Siegert states 
for $l>0$ and for long-range potentials but the algebraic algorithm 
developed in Ref.~\cite{TON98} must be replaced, most probably 
by some iterative technique. 

We have shown that, unexpectedly, the method of Ref.~\cite{TON98} 
can also be used to construct scattering wave functions 
for potentials with long-range terms such as the centrifugal term. 
However, in such cases, the $R$-matrix method of Refs.~\cite{BHS98,HSV98} 
is more advantageous since (i) it is simpler, (ii) it also provides 
the $S$ matrix, and (iii) it is readily extended to the multichannel case. 
\section*{Acknowledgements}
This text presents research results of the Belgian program P4/18 on 
interuniversity attraction poles initiated by the Belgian-state 
Federal Services for Scientific, Technical and Cultural Affairs.
J.-M.S. is supported by the Fonds National de la Recherche Scientifique 
of Belgium. 
\setcounter{equation}{0}

\section*{Appendix A}
Let $\ve{B}$ be an invertible matrix and $u$ and $v$ be vectors. 
The inverse of the matrix 
\beq
\ve{A} = \ve{B} + uv^T
\eeqn{A.1}
is given by
\beq
\ve{A}^{-1} = \ve{B}^{-1} - \frac{\ve{B}^{-1} uv^T \ve{B}^{-1}}
{1 + v^T \ve{B}^{-1} u},
\eeqn{A.2}
where the denominator is a scalar.
A corollary of Eq.~(\ref{A.2}) reads 
\beq
\ve{A}^{-1} u = \frac{\ve{B}^{-1} u}{1 + v^T \ve{B}^{-1} u}.
\eeqn{A.3a}
Another corollary is the relation 
\beq
(v^T \ve{A}^{-1} u)^{-1} = 1 + (v^T \ve{B}^{-1} u)^{-1}
\eeqn{A.3}
from which Eq.~(\ref{2.13}) follows. 

The norm matrix (\ref{3.6}) has the form 
\beq
\ve{N} = \ve{1} + \alpha uu^T.
\eeqn{A.4}
where $u$ is here a unit vector ($||u||=1$).
Arbitrary powers of $\ve{N}$ are given by 
\beq
\ve{N}^{\lambda} = \ve{1} + [(1+\alpha)^{\lambda} - 1] uu^T,
\eeqn{A.5}
for any $\lambda$, integer or fractional, positive or negative. 
In Eq.~(\ref{3.6}), the components of the unit vector $u$ read 
\beq
u_i = (-1)^i \frac{1}{N} \sqrt \frac{1-x_i}{x_i}
\eeqn{A.6}
and the coefficent $\alpha$ is given by 
\beq
\alpha = N^2/(2N+1).
\eeqn{A.7}
\newpage
\begin{figure}[ht]
\includegraphics{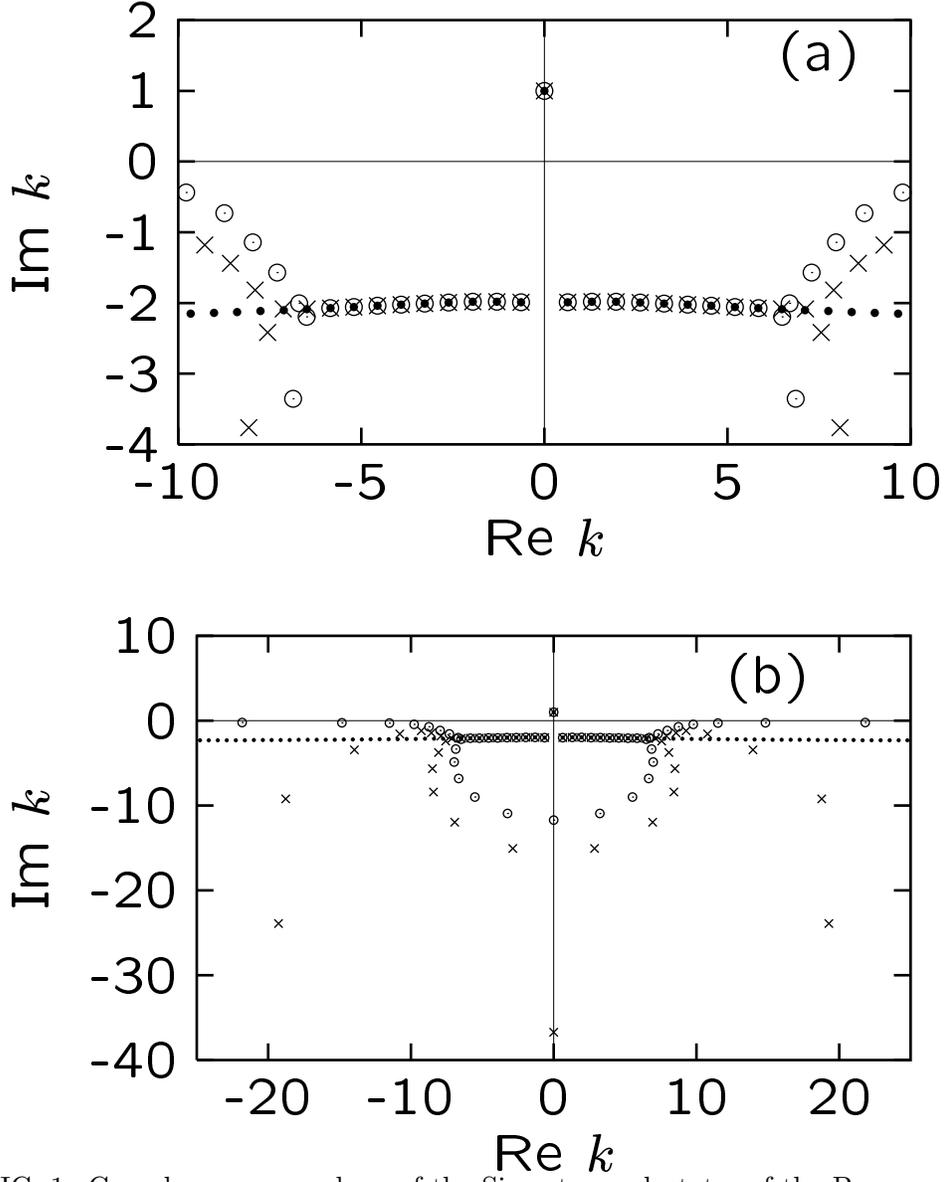}
\caption{Complex wave numbers of the Siegert pseudostates 
of the Bargmann potential: 
exact (dots), without (circles) and with (crosses) Gauss 
approximation for the overlap matrix 
($a = 5$ and $N = 25$).}
\label{fig:1}
\end{figure}
\begin{table}[ht]
\caption{Symbols of Ref.~\protect\cite{TON98} expressed 
in the present notations ($l=0$).
\label{tab:1}}
\begin{center}
\begin{tabular}{cccc}
Ref.~\protect\cite{TON98} & Eq. & Present & Eq. \\
\hline
 $x_i$ & (C8) & $2x_i-1$ & (\protect\ref{3.2}) \\
 $\kappa_i$  & (C11) & 2$\lambda_i$ & (\protect\ref{3.5}) \\
 $\pi_i (x)$ & (C9) & $(8a)^{1/2} x_i (x+1)^{-1} f_i [\dem a(x+1)]$ 
 & (\protect\ref{3.3}) \\
 $\rho_{ij}$ & (C19) & 
 $a^2 x_i x_j \la f_i | f_j \ra \equiv a^2 x_i x_j N_{ij}$ 
 & (\protect\ref{3.6}) \\
 $\tilde{K}_{ij}$ & (C20) & $a^2 x_i x_j \la f_i | T_0 + \cL(0) | f_j \ra$ 
 & (\protect\ref{3.9}),(\protect\ref{3.10})  \\
 $L_{ij}$ & (C18) & 
 \dem $a^2 x_i x_j \la f_i | r^{-1} \delta (r-a) | f_j \ra$ & 
 (\protect\ref{3.12}) \\
 $\ve{\widetilde{H}} + (1-ika) \ve{L}$ & (C15) & $a^2 x_i x_j \ve{C}(ika)$ 
 & (\protect\ref{3.1}) 
\end{tabular}
\end{center}
\end{table}
\begin{table}[ht]
\caption{Complex wave numbers of the Siegert pseudostates 
of the Bargmann potential: exact [Eq.~(\protect\ref{5.5})], 
without [Eq.~(\protect\ref{3.6})]
and with [Eq.~(\protect\ref{3.7})] Gauss approximation  
for the overlap matrix ($a = 5$ and $N = 25$). 
\label{tab:2}}
\begin{center}
\begin{tabular}{cccccc}
\multicolumn{2}{c}{Exact} &
\multicolumn{2}{c}{Eq.~(\protect\ref{3.6})} & 
\multicolumn{2}{c}{Eq.~(\protect\ref{3.7})} \\ 
\hline
Re\ $k_0^{(n)}$ & Im\ $k_0^{(n)}$ & Re\ $k_0^{(n)}$ & Im\ $k_0^{(n)}$ 
& Re\ $k_0^{(n)}$ & Im\ $k_0^{(n)}$ \\
\hline
 0   	      & 0.9999999999955 & 0.0000000 & $ 0.9999999$ &  0.0000000 & $ 0.9999999$ \\
 0.6390266702 & $-1.9909403640$ & 0.6390286 & $-1.9909403$ &  0.6390286 & $-1.9909403$ \\
 1.2947020189 & $-1.9808654900$ & 1.2947061 & $-1.9808653$ &  1.2947061 & $-1.9808653$ \\
 1.9559919327 & $-1.9824365206$ & 1.9559981 & $-1.9824363$ &  1.9559981 & $-1.9824363$ \\
 2.6142071140 & $-1.9926145823$ & 2.6142153 & $-1.9926144$ &  2.6142153 & $-1.9926144$ \\
 3.2675033086 & $-2.0070370919$ & 3.2675133 & $-2.0070370$ &  3.2675135 & $-2.0070369$ \\
 3.9163165163 & $-2.0231436137$ & 3.9163283 & $-2.0231433$ &  3.9163285 & $-2.0231433$ \\
 4.5614908509 & $-2.0396289360$ & 4.5615045 & $-2.0396283$ &  4.5615048 & $-2.0396285$ \\
 5.2037975084 & $-2.0558538335$ & 5.2038193 & $-2.0558336$ &  5.2038125 & $-2.0558513$ \\
 5.8438509338 & $-2.0715187039$ & 5.8453967 & $-2.0710367$ &  5.8438594 & $-2.0715509$ \\
 6.4821222431 & $-2.0864975209$ & 6.4948355 & $-2.1966169$ &  6.4817804 & $-2.0855307$ \\
 7.1189698354 & $-2.1007533271$ & 6.6971747 & $-2.0045040$ &  7.1321815 & $-2.0820124$ \\
 7.7546673954 & $-2.1142944575$ & 7.2971683 & $-1.5698719$ &  7.5590343 & $-2.4165539$ \\
\end{tabular}		     
\end{center}
\end{table}
\begin{table}[ht]
\caption{Phase shifts (in degrees) of the Bargmann potential: 
exact [Eq.~(\protect\ref{5.3})], 
Siegert pseudostate method without [Eq.~(\protect\ref{3.6})] 
and with [Eq.~(\protect\ref{3.7})] Gauss approximation  
for the overlap matrix, 
and $R$ matrix [Eq.~(\protect\ref{2.10})]. 
\label{tab:3}}
\begin{center}
\begin{tabular}{ccccccc}
$E$ & Exact & $a$ & $N$ & Eq.~(\protect\ref{3.6}) 
& Eq.~(\protect\ref{3.7}) & Eq.~(\protect\ref{2.10}) \\
\hline
 0.1 & 143.30077480 & 5 & 25 & 143.30076489 & 143.30076489 & 143.30076490 \\
     &              &   & 40 & 143.30076866 & 143.30076866 & 143.30076866 \\
     &              & 6 & 25 & 143.30079803 & 143.30079803 & 143.30079803 \\
     &              &   & 40 & 143.30077472 & 143.30077472 & 143.30077473 \\
  1  &  90.00000000 & 5 & 25 &  89.99999083 &  89.99999083 &  89.99999083 \\
     &              &   & 40 &  89.99999947 &  89.99999947 &  89.99999958 \\
     &              & 6 & 25 &  90.00005685 &  90.00005678 &  90.00005688 \\
     &              &   & 40 &  89.99999998 &  89.99999998 &  90.00000007 \\
  10 &  36.69922520 & 5 & 25 &  36.69916732 &  36.69916703 &  36.69916730 \\
     &              &   & 40 &  36.69922475 &  36.69922475 &  36.69922502 \\
     &              & 6 & 25 &  36.69975390 &  36.69974764 &  36.69974770 \\
     &              &   & 40 &  36.69922520 &  36.69922520 &  36.69922526
\end{tabular}		     
\end{center}
\end{table}
\end{document}